\documentclass[cernpreprint,atlasdraft=true,texlive=2020,UKenglish,texmf]{atlasdoc}
 
\usepackage{atlaspackage}
\usepackage{atlasbiblatex}
 
\usepackage{atlasphysics}
\usepackage{multirow}
 
\usepackage{listings}
 
\usepackage{color}
 
\definecolor{dkgreen}{rgb}{0,0.6,0}
\definecolor{gray}{rgb}{0.5,0.5,0.5}
\definecolor{mauve}{rgb}{0.58,0,0.82}
 
\graphicspath{{logos/}{figures/}}
 
\usepackage{ANA-SUSY-2018-14-PAPER-defs}

\AtlasTitle{Search for displaced leptons in $\sqs = 13$~\TeV\ \pp\ collisions with the ATLAS detector}
 
\AtlasVersion{4.0}
\AtlasAbstract{
A search for charged leptons with large impact parameters using 139~\ifb\ of $\sqrt{s} = 13$~\TeV\ \pp\ collision data from the ATLAS detector at the LHC is presented, addressing a long-standing gap in coverage of possible new physics signatures. Results are consistent with the background prediction. This search provides unique sensitivity to long-lived scalar supersymmetric lepton-partners (sleptons). For lifetimes of 0.1~ns, selectron, smuon and stau masses up to 720~\GeV, 680~\GeV, and 340~\GeV~are respectively excluded at 95\% confidence level, drastically improving on the previous best limits from LEP.
 
}
 
\author{The ATLAS Collaboration}
 
\AtlasRefCode{SUSY-2018-14}

\PreprintIdNumber{CERN-EP-2020-205}
 
\AtlasJournalRef{Physical Review Letters 127, 051802 (2021)}
\AtlasDOI{10.1103/PhysRevLett.127.051802}

\hypersetup{pdftitle={ANA-SUSY-2018-14},pdfauthor={The ATLAS Collaboration}}
 
\begin{document}
 
\maketitle

Particles with long lifetimes are a feature of the Standard Model (SM) and many theories beyond the Standard Model (BSM) including $R$-parity-conserving supersymmetry (SUSY)~\cite{Farrar:1978xj,Golfand:1971iw,Volkov:1973ix,Wess:1974tw,Wess:1974jb,Ferrara:1974pu,Salam:1974ig}
models like split-SUSY~\cite{Giudice:2004tc,ArkaniHamed:2004fb} and
gauge-mediated SUSY breaking (GMSB)~\cite{Dine:1981gu,AlvarezGaume:1981wy,Nappi:1982hm}, as well as $R$-parity-violating SUSY models~\cite{Barbier:2004ez,Allanach:2006st}
and exotic scenarios such as universal extra dimensions~\cite{Appelquist:2000nn,Cheng:2002ab}. However, particle lifetime remains an underexplored parameter of phase space at the Large Hadron Collider (LHC), where detectors and searches for new physics were designed to measure the decay products of short-lived, heavy particles with the assumption that those decay products trace back to the collision point, or very close to it~\cite{SUSY-2011-23,SUSY-2013-11,SUSY-2018-32,CMS-SUS-13-006,CMS-SUS-17-009}.
BSM particles with lifetimes longer than a few picoseconds produce unconventional signatures, including displaced decay products that do not trace back to the interaction point. This brings technical challenges in almost all aspects of the search; consequently, some models with \TeV-scale BSM particles in this lifetime regime remain unexplored.
While many dedicated searches for long-lived particles have been performed by the ATLAS~\cite{EXOT-2018-61,Aad:2020srt,SUSY-2017-04,EXOT-2017-25,EXOT-2017-05,EXOT-2017-03,EXOT-2017-26,SUSY-2016-08,SUSY-2016-06,SUSY-2016-03,SUSY-2014-09,SUSY-2014-02,SUSY-2014-09,EXOT-2013-12} and CMS~\cite{,Sirunyan:2020cao,CMS-PAS-EXO-19-013,CMS-SUS-14-020,CMS-SUS-19-005,CMS-EXO-17-018,Sirunyan:2019wau,Sirunyan:2019gut, Sirunyan:2019wau,Sirunyan:2018vlw,Sirunyan:2018ldc,Sirunyan:2018pwn,CMS-PAS-EXO-16-022,CMS:2014hka} Collaborations, signatures with displaced leptons with no visible decay vertex would not be identified by any previous ATLAS search.
This Letter addresses that gap in coverage.
 
This signature brings unique sensitivity to GMSB SUSY models~\cite{Alwall:2008ve,Alwall:2008ag,Alves:2011wf}, where the nearly massless gravitino is the lightest SUSY particle (LSP), and the next-to-lightest SUSY particle (NLSP) becomes long-lived due to the small gravitational coupling to the LSP. Well-motivated versions of this model have a stau ($\tilde{\tau}$) as the single NLSP, or a selectron ($\tilde{e}$), smuon ($\tilde{\mu}$), and $\tilde{\tau}$ as co-NLSPs~\cite{Evans_2016}. In these models, pair-produced sleptons ($\tilde{\ell}$) of the same flavor decay into an invisible gravitino and a charged lepton of the same flavor as the parent $\tilde{\ell}$.
A combination of results from the LEP experiments excluded the superpartners of the right-handed muons and electrons ($\tilde{\mu}_R$ and $\tilde{e}_R$) of any lifetime for masses less than 96.3~\GeV\ and 65.8~\GeV. The OPAL experiment alone set the best limits for all lifetimes of $\tilde{\tau}_1$, a mixture of the superpartners of the left- and right-handed $\tau$-leptons, and excluded masses less than 87.6~\GeV~\cite{Heister:2002vh,Abbiendi:2005gc,L3,Abdallah:2002rd,LEPlimitsSlepton}.
A previous search from the CMS experiment~\cite{CMS-B2G-12-024} selected events with displaced, different-flavor leptons using 19.7~\ifb\ of 8~\tev\ data, but did not directly target $\tilde{\ell}$ decays.
A reinterpretation concluded that OPAL's constraints remained the most stringent~\cite{Evans_2016}. Additionally, Ref.~\cite{Cahill-Rowley:2015dyd} shows that targeting this signature could help improve the coverage of minimal supersymmetric models with a gravitino LSP.
The present search provides mass sensitivity beyond the LEP limits.
 
To evaluate signal sensitivity, Monte Carlo (MC) events in a simplified GMSB SUSY model were simulated with up to two additional partons at leading order using \MGMCatNLOV{v2.6.1}~\cite{Alwall:2014hca} with the \nnpdftwo parton distribution function (PDF) set~\cite{Ball:2012cx}, and interfaced to \PYTHIAV{8.230}~\cite{Sjostrand:2006za} using the A14 set of tuned parameters (tune)~\cite{ATL-PHYS-PUB-2014-021}. The sparticle decay was simulated using $\GEANT4$~\cite{Agostinelli:2002hh}, which does not preserve information about the chirality of the $\tilde{\ell}$. The mixed states of the superpartners of the left- and right-handed $\tau$-leptons, $\tilde{\tau}_{1,2}$, were generated with mixing angle $\sin\theta_{\tilde\tau}=0.95$. The impact of multiple interactions in the same and neighboring bunch crossings (pileup) was modeled by overlaying each hard-scattering event with simulated minimum-bias events generated with \PYTHIAV{8.210}~\cite{Sjostrand:2006za} using the A3 tune~\cite{ATL-PHYS-PUB-2016-017} and \nnpdftwo PDF set~\cite{Ball:2012cx}. Signal cross sections were calculated at next-to-leading order in $\alpha_\text{s}$, with soft-gluon emission effects added at next-to-leading-logarithm accuracy ~\cite{Beenakker:1999xh,Debove:2010kf,Fuks:2012qx,Fuks:2013vua,Fiaschi:2018hgm}. The nominal cross section and uncertainty were taken from an envelope of predictions using different PDF sets and factorization and renormalization scales~\cite{Borschensky:2014cia}. The simplified model used for interpretation assumes the superpartners of the left- and right-handed leptons are mass degenerate, yielding a cross section of $0.37 \pm 0.01$~pb for a single flavor of $\tilde{\ell}$ with mass $100~\GeV$ and $0.059 \pm 0.004$~fb for a $\tilde{\ell}$ with mass $800~\GeV$. Simulated events were generated for $\tilde{e}$/$\tilde{\mu}$ ($\tilde{\tau}$) masses 50--900~\GeV\ (50--400~\GeV) and lifetimes 0.01--10~ns (0.1--1~ns).

This search uses $139~\ifb$ of data collected by the ATLAS experiment from \pp collisions at $\sqrt{s} = 13~\TeV$. The ATLAS detector consists of concentric subdetectors used together to identify particles\footnote{ATLAS uses a right-handed coordinate system with its origin at the nominal interaction point (IP) in the center of the detector and the $z$-axis along the beam pipe. The $x$-axis points from the IP to the center of the LHC ring, and the $y$-axis points upward. Cylindrical coordinates $(r,\phi)$ are used in the transverse plane, $\phi$ being the azimuthal angle around the $z$-axis. The pseudorapidity is defined in terms of the polar angle $\theta$ as $\eta=-\ln\tan(\theta/2)$.}~\cite{PERF-2007-01,ATLAS-TDR-2010-19,PIX-2018-001}. Data collection relies on a two-level trigger system, which uses tracking information from the inner detector (ID) along with information from the calorimeters and muon spectrometer (MS) to make fast, event-level decisions~\cite{TRIG-2016-01}. The typical lepton selection algorithms used in the trigger select particles coming from the primary interaction and cannot be used to select displaced leptons. Instead, triggers without tracking information are used: electrons are identified using only their electromagnetic calorimeter (EM) signature via photon triggers, and muons are identified using MS information only. Single-photon and diphoton triggers select EM signatures with energy greater than 140~\GeV\ and 50~\GeV, respectively, and the muon trigger selects MS signatures with transverse momentum (\pt) greater than 60~\GeV\ in the range $|\eta| < 1.05$. These triggers have an acceptance independent of lepton displacement in the range probed by this search. The acceptance ranges from 1\% to 80\% for all flavors, increasing with $\tilde{\ell}$ mass, and is lower for $\tilde{\tau}$ than $\tilde{e}$ or $\tilde{\mu}$ due to the smaller \pt of the final-state leptons.

After the trigger stage, more complex tracking algorithms are possible, and tracks can be used more extensively for particle identification.
Displaced leptons are identified as those with large transverse impact parameter ($\absdz$), the distance of closest approach of the particle's track to the interaction point in the $x$--$y$ plane. The \absdz is measured relative to the vertex with the highest $\Sigma \pt^{2}$ of associated tracks. Tracks are reconstructed by fitting a series of ID hits to identify those consistent with a particle's trajectory. For this search, tracking is performed in two stages: first, standard tracking reconstructs tracks with $\absdz < 10$~mm~\cite{PERF-2015-08}, then an additional reconstruction step uses hits not matched to tracks in the previous stage, adding tracks with $\absdz < 300$~mm~\cite{ATL-PHYS-PUB-2017-014}.
The extended track collection is combined with EM energy clusters to reconstruct electrons, or with tracks composed of segments measured in the MS to reconstruct muons, both in the range $|\eta| < 2.5$. Standard lepton identification algorithms~\cite{PERF-2015-10,EGAM-2018-01,Aad:2020gmm} are modified by removing requirements on \absdz and the number of hits matched to the track. Figure~\ref{fig:lep_eff} shows the final reconstruction efficiency for displaced electrons and muons.

\begin{figure}[htbp]
\centering
\includegraphics[width=.9\textwidth]{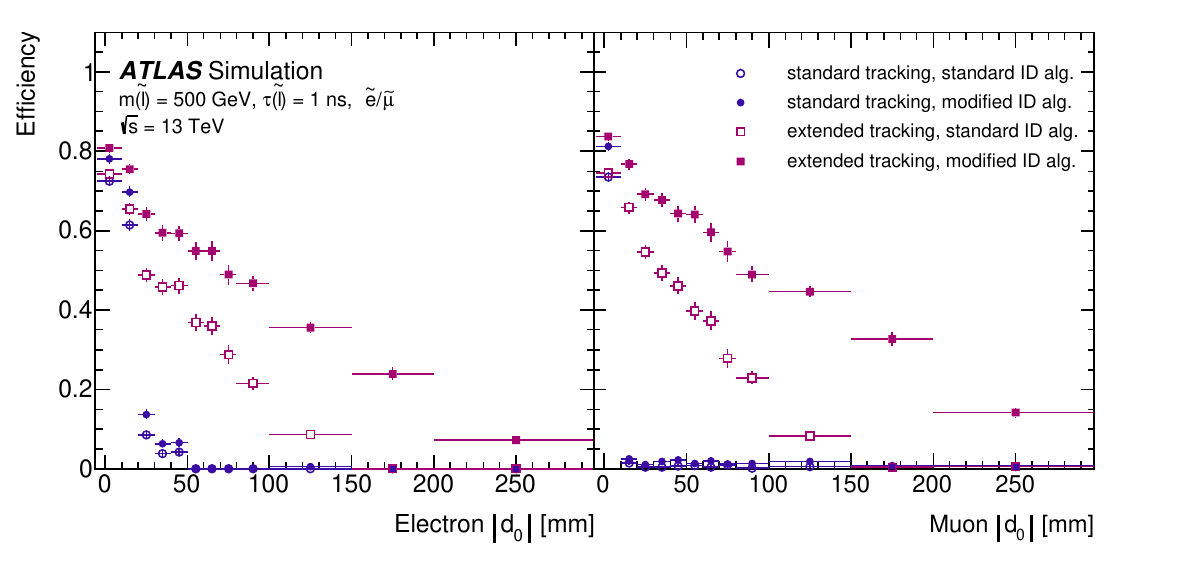}
\caption{Electron (left) and muon (right) reconstruction and identification efficiency in signal MC simulation. Leptons result from the decay of a $\tilde{\ell}$ with $m_{\tilde{\ell}} = 500~\GeV$ and $\tau_{\tilde{\ell}} = 1$~ns. Efficiency is defined as the number of reconstructed leptons divided by the number of generator-level leptons. Both the reconstructed and generator-level leptons are required to have $\pt > 20$~\GeV\ and $|\eta| < 2.5$. The closed purple square markers show the final lepton reconstruction efficiency. Markers are placed at the bin centers.}
\label{fig:lep_eff}
\end{figure}
 
Signal leptons must have high transverse momentum, $\pt > 65~\gev$, and large transverse impact parameter, $3~\text{mm} < \absdz < 300$~mm, to remove SM backgrounds. To reduce the background from out-of-time cosmic-ray muons, a requirement is placed on the MS timing relative to when a standard model particle is expected to arrive in the detector ($t_{0}$). The average time measured by the muon's MS track segments, $t_{0}^\text{avg}$, must have an absolute value less than $30$~ns.
In order to reduce the contribution from leptons from decays of heavy-flavor hadrons, signal leptons are required to be isolated from nearby activity in the ID and calorimeters. The sum of the \pt of all tracks near an electron (muon) must be less than 6\% (4\%) of the lepton \pt, and the sum of energy deposits near the electron (muon) in the calorimeters must be less than 6\% (15\%) of the lepton's energy~\cite{PERF-2015-10,EGAM-2018-01}.
The remaining quality criteria are used to minimize backgrounds and are inverted in the data-driven background estimation. Signal leptons must satisfy these to remove fake leptons originating from the mismatching of ID tracks to MS tracks or to calorimeter signatures. ID tracks associated with leptons are required to have a fit with $\chi^2/n_{\mathrm{DOF}} < 2$ and no more than one missing hit after their innermost hit. Consistency between the two components of the reconstructed lepton is required.
For electrons, this is ensured by requiring the ID track \pt measurement to be no less than half the electron \pt measured when accounting for the calorimeter energy ($(\pt^\text{track}-\pt^e)/\pt^e > -0.5$), and the combined fit of the muon's ID and MS tracks must satisfy $\chi^2/n_{\mathrm{DOF}} < 3$.  Muons are required to have measurements in at least three precision tracking layers of the MS and at least one high-precision $\phi$ measurement.
 
Three orthogonal signal regions are defined with at least two signal leptons and are distinguished by the flavor of the two highest-\pt leptons: SR-$ee$ with two electrons, SR-$\mu\mu$ with two muons, and SR-$e\mu$ with one muon and one electron. No requirements are placed on the charge of the leptons. In order to ensure the broad applicability of this result to other models, event-level requirements beyond the presence of the two signal leptons are minimal.
Backgrounds from lepton pairs produced via interactions with detector material are reduced by requiring that the opening angle between the two leptons, $\Delta R_{\ell\ell} \equiv \sqrt{(\Delta\eta_{\ell\ell})^{2} + (\Delta \phi_{\ell\ell})^{2}}$, is greater than 0.2.
Additionally, the event must not contain any cosmic-tagged muons. A cosmic-ray muon traversing the detector coincident with an LHC collision leaves a signature that could be reconstructed as two back-to-back muons, one in the top half of the detector, $\mu_t$ and the other in the bottom, $\mu_b$.
Each muon is tagged as resulting from a cosmic-ray muon if it has MS segments along its trajectory on the opposite side of the detector, or if its trajectory traces back to a gap in detector coverage~\cite{Aad:2020srt}. A window in $\eta$ and $\phi$ is defined relative to the muon's trajectory, and if an MS segment is found within $|\eta_\mu + \eta_\text{MS segment}|  < 0.018 \textrm{ and } |(\phi_\mu - \phi_\text{MS segment}) - \pi| < 0.25$, the muon is cosmic-tagged. This algorithm has a cosmic rejection efficiency of $> 99\%$.
 
The number of background events remaining after signal selections is estimated from data while keeping the signal regions blinded. In SR-$ee$ and SR-$e\mu$, the dominant background comes from fake leptons, with a smaller contribution from leptons from heavy-flavor hadron decays. Zero events with a cosmic-tagged muon and electron were observed; therefore, the background contribution from untagged cosmic-ray muons in SR-$e\mu$ is expected to be negligible. Fake electrons typically result from the mismatching of a track to a photon.
Fake muons result from the mismatching of an ID track to an MS track, and are comparatively rare since there is less activity and better pointing information in the MS than in the calorimeter.
Fake leptons tend to fail quality criteria, as a result, they have poor $\chi^2$ or inconsistent track and lepton $\pt$. Moreover, these requirements also remove heavy-flavor contributions which tend to have extra energy in their clusters compared to their tracks. As a result, the contribution of these backgrounds is estimated together.
The quality criteria in this analysis are uncorrelated between the two leptons in an event, which has been verified in inverted regions in data. Since the variables are uncorrelated, they can be used to estimate the background contribution to the signal regions. The background is estimated with an ABCD method~\cite{Kasieczka:2020pil} by calculating the ratio of the number of events where lepton 1 passes inverted quality criteria (not including lepton $\pt$ or $\absdz$) and lepton 2 passes nominal requirements, and vice versa, divided by the number of events where both leptons fail the quality criteria. To estimate the background in SR-$ee$, where the two leading leptons are electrons, lepton 1 is the leading electron, and lepton 2 is the subleading electron. To estimate the background in SR-$e\mu$, where the two leading leptons are an electron and a muon, lepton 1 and 2 are the leading electron and muon, respectively. The same algorithm is used for SR-$ee$ and SR-$e\mu$, but due to statistical limitations in SR-$e\mu$, the \pt and \absdz requirements on the leptons are relaxed to $\pt > 50~\gev$ and $\absdz > 2$ mm. As the \pt and \absdz distributions are exponentially falling, this results in a conservative background estimate in SR-$e\mu$.
 
In the ABCD method, the phase space is split into 4 regions: region A, region B, region C, and region D. Region A is the signal region, where all requirements are satisfied, region B is the region where lepton 1 fails quality criteria but lepton 2 passes all lepton requirements, region C is the region where lepton 2 fails quality criteria but lepton 1 passes all requirements, and region D is the region where both leptons fail quality criteria. For an electron, the inverted quality criteria are: ID track $\chi^2/n_{\mathrm{DOF}} > 2$, $(\pt^\text{track}-\pt^e)/\pt^e < -0.5$, and greater than one missing hit after the electron's innermost hit. For a muon, the inverted quality criteria are: ID track $\chi^2/n_{\mathrm{DOF}} > 2$, combined MS and ID track $\chi^2/n_{\mathrm{DOF}} > 3$, measurements in less than three precision tracking layers of the MS, greater than one missing hit after the muon's innermost hit, and no high-precision $\phi$ measurement.
The number of events in the signal region is then estimated by the following calculation,
\begin{equation*}
N_{A}^\text{predicted} = \frac{N_{B}\times N_{C}}{N_D},
\end{equation*}
where $N_A^\text{predicted}$ is the predicted number of background events in the signal region (region A), $N_B$ is the number of events in region B, $N_C$ is the number of events in region C, and $N_D$ is the number of events in region D.
 
\begin{table}[htb]
\centering
\caption{Validation of the background data-driven estimate for the $ee$ and $e\mu$ channel fake and heavy-flavor backgrounds.  Uncertainties are statistical.}
\vspace{0.2cm}
\begin{tabular}{l   c  c c c }
\hline
&\multicolumn{1}{c}{VR-$ee$-fake} & \multicolumn{1}{c}{VR-$ee$-heavy-flavor}  & \multicolumn{1}{c}{VR-$e\mu$-fake} & \multicolumn{1}{c}{VR-$e\mu$-heavy-flavor}\\
\hline
Estimate & $1356 \pm 49$ & $23.5 \pm 1.9$ & $1.9^{+1.8}_{-1.0}$ &  $0.38^{+0.37}_{-0.32}$\\
Observed & 1440 & 26& 2 & 1\\
\hline
\end{tabular}
\label{tab:val_fakes}
\end{table}
 
Validations of these background estimates are performed, with the heavy-flavor and fake contributions targeted separately. The validation of the heavy-flavor contribution is achieved using the same method as the nominal background estimation but inverting the isolation requirement in all regions. To increase statistics, the requirement on $(\pt^\text{track}-\pt^e)/\pt^e$ is loosened to be greater than $-0.9$ instead of $-0.5$, as this distribution exponentially decreases from -1 to -0.5. The fake-lepton contribution is probed by inverting the most powerful fake discriminators by requiring the electron variable, $(\pt^\text{track}-\pt^e)/\pt^e$, to be less than $-0.5$, and the muon's combined track's $\chi^2/n_{\mathrm{DOF}}$ to be greater than $3$, and performing the ABCD estimate with the remaining quality criteria.
The validation of both estimates is shown in Table~\ref{tab:val_fakes}. Even with the loosened requirements of $\pt > 50~\gev$ and $\absdz > 2$ mm in VR-$e\mu$-fake and VR-$e\mu$-heavy-flavor, and $(\pt^\text{track}-\pt^e)/\pt^e > -0.9$ in VR-$e\mu$-heavy-flavor, the statistics in these validation regions are limited. The background is small since fake muons are rare, and the requirements on $\pt$ and $\absdz$ on signal leptons render heavy-flavor backgrounds negligible. Nonetheless, the numbers of estimated and observed events were consistent within statistical uncertainties, and uncertainties were assigned to account for small differences between predictions and observations in each validation. The predicted number of background events from fake and heavy-flavor-decay leptons is $0.46 \pm 0.10$ in SR-$ee$ and $0.007^{+0.019}_{-0.007}$ in SR-$e\mu$, including all uncertainties.
 
The dominant background in SR-$\mu\mu$ comes from mis-measured reconstructed muons from cosmic rays. The fake lepton background is found to be negligible due to the rarity of fake muons. The heavy-flavor background is estimated using an ABCD estimate extrapolating from non-isolated muons to isolated muons with loosened $\pt$ and $\absdz$ requirements to increase statistics ($\
pt > 50$ GeV and subleading muon $\absdz > 0.5$ mm). This results in a heavy-flavor estimate of $< 10^{-4}$ events. For a cosmic event to be a background to this search, both $\mu_t$ and $\mu_b$ must be reconstructed in the same event, which means their $|t_0^{\mathrm{avg}}|$ will be near the edges of the allowed range and are likely to have their MS hits associated with the wrong event. This results in reconstructed muons with good quality ID tracks, but poor quality MS signatures, which could present challenges in cosmic-tagging one or both muons. An event with a cosmic-ray muon could meet signal region requirements if both muons have missing MS hits and neither is tagged. Cosmic-tagging failures occur not when the muon in question is mis-measured, but when the muon is in the half of the detector opposite to a poorly reconstructed MS track, and no MS segments are found in the tag window. The estimate of this background relies on the assumption that the quality of a muon and its probability to be cosmic-tagged are uncorrelated.
 
All events considered in this estimate have $\mu_b$ passing all signal requirements, while $\mu_t$ is either cosmic-tagged, fails to satisfy some of the quality criteria, or both. No dimuon events were observed with two muons on the same side of the detector. In events where $\mu_t$ is cosmic-tagged, the ratio of $\mu_t$ which satisfy the quality criteria to those that do not, $R_{\mathrm{good}}$, is measured. This ratio is multiplied by the number of events in which $\mu_t$ is not cosmic-tagged, but fails to satisfy at least one of the quality criteria, to estimate the background in SR-$\mu\mu$. The estimate is validated by redefining the cosmic-tag window to leave more muons untagged, providing a larger sample for studying $R_{\mathrm{good}}$. An additional uncertainty is assigned to the background estimate from the validation to account for the \absdz dependence of $R_{\mathrm{good}}$, which cannot be directly constrained in the nominal estimate due to statistical limitations. Additional validations test other assumptions by varying the quality criteria and reversing the roles of $\mu_b$ and $\mu_t$ in the definition of $R_{\mathrm{good}}$. Including all uncertainties, $0.11^{+0.20}_{-0.11}$ events are predicted in SR-$\mu\mu$.
 
Signal systematic uncertainties are evaluated to quantify differences between data and simulation and correct the MC events where possible. Differences in signal lepton selection efficiency cannot be compared between data and MC simulation due to the lack of displaced leptons in data, so a conservative systematic uncertainty is derived in three steps. First, trigger, reconstruction, and selection efficiencies are measured for low-\absdz leptons resulting from $Z$ boson decays, for which data and simulation can be compared. Scale factors are derived to correct the simulation to match the data. Uncertainties in these scale factors are statistical and less than 5\%. Next, the high-\absdz tracking efficiency is compared between signal simulation and data with cosmic-tagged muons. After corrections to account for the different physical processes, the tracking efficiency as a function of displacement is compared and an 8\% uncertainty is assigned to each lepton. Finally, the \absdz dependence of the lepton reconstruction and selection efficiency is compared with the \absdz dependence of the tracking efficiency in simulation only. The variation of the selection efficiency as a function of \absdz is taken as an uncertainty to account for any discrepancies that cannot be studied in data. This uncertainty increases with displacement and is 0.5--5\% (3--27\%) for muons (electrons). It is larger for electrons due to identification challenges introduced by the ambiguity in the detector signatures of electrons, photons, and converted photons. Theoretical uncertainties include cross section uncertainties of 2--6\%, and effects of varying the factorization and renormalization scales, $< 5$\%. Other uncertainties, including the impact of pileup on signal selection, luminosity uncertainty~\cite{ATLAS-CONF-2019-021,LUCID2}, and uncertainty from the filtering selection used for the extended track reconstruction, contribute at $< 2\%$.
 
\begin{table}[ht!]
\centering
\caption{The expected and observed yields in the signal regions. Combined statistical and systematic uncertainties are presented. Estimates are truncated at 0 if the size of measured systematic uncertainties would yield a negative result.}
\vspace{0.2cm}
\begin{tabular}{lccc}
\hline
Region 			& SR-$ee$ 			& SR-$\mu\mu$ 				& SR-$e\mu$ \\
\hline
\hline
\\[-1em]
Fake + heavy-flavor 			& $0.46 \pm 0.10$ 	& $< 10^{-4}$						& $0.007^{+0.019}_{-0.007}$\\
\\[-1em]
Cosmic-ray muons 		& --				& $0.11 ^{+0.20}_{-0.11}$ 	& -- \\
\\[-1em]
\hline
\\[-1em]
Expected background 	& $0.46 \pm 0.10$ 	& $0.11 ^{+0.20}_{-0.11}$	& $0.007^{+0.019}_{-0.007}$\\
\\[-1em]
\hline
\hline
\\[-1em]
Observed events & 0				& 0 						& 0 \\
\\[-1em]
\hline
\end{tabular}
\label{tab:results_yields}
 
\end{table}
 
Zero events are observed in each of the three signal regions, consistent with the background predictions shown in Table~\ref{tab:results_yields}. As no excess of events is observed, exclusion limits on the $\tilde{\ell}$ masses are derived at 95\% confidence level (CL) following the CL$_\text{s}$ prescription~\cite{Read:2002hq}. The HistFitter package~\cite{Baak:2014wma} is used for statistical interpretation, and all systematic uncertainties are treated as Gaussian nuisance parameters during the fitting procedure.  SR-$ee$ and SR-$\mu\mu$ are fit individually to calculate limits on GMSB SUSY models with a $\tilde{e}$ or $\tilde{\mu}$ NLSP, while $\tilde{\tau}$ NLSP and co-NLSP limits are obtained using a simultaneous fit of all three signal regions. All uncertainties other than statistical are treated as correlated across the orthogonal regions.

Limits on long-lived $\tilde{\ell}$ production are presented in Figure~\ref{fig:excl_limit}, where expected and observed exclusion contours as a function of $\tilde{\ell}$ mass and lifetime are shown.
For a lifetime of 0.1~ns, $\tilde{e}$ NLSP, $\tilde{\mu}$ NLSP, $\tilde{\tau}$ NLSP, and co-NLSP scenarios are excluded for $\tilde{\ell}$ masses up to 720~\GeV, 680~\GeV, 340~\GeV, and 820~\GeV, respectively, for the case where the superpartners of the left- and right-handed leptons are mass degenerate. For a direct comparison with the previous best limits available from LEP, superpartners of right-handed electrons ($\tilde{e}_R$), muons ($\tilde{\mu}_R$), and left-handed $\tau$-leptons ($\tilde{\tau}_L$) are excluded up to 580~\GeV, 550~\GeV, and 280~\GeV, respectively, for lifetimes of 0.1~ns. This result probes GMSB $\tilde{\ell}$ production for the first time in this lifetime range at the electroweak scale and approaching the \TeV\ scale. Furthermore, as no requirements were made on missing energy, displaced vertices, or jets, this result is model-independent and applicable to any BSM model producing high-\pt displaced leptons.
 
\begin{figure}[h!]
\centering
\includegraphics[width=0.6\textwidth]{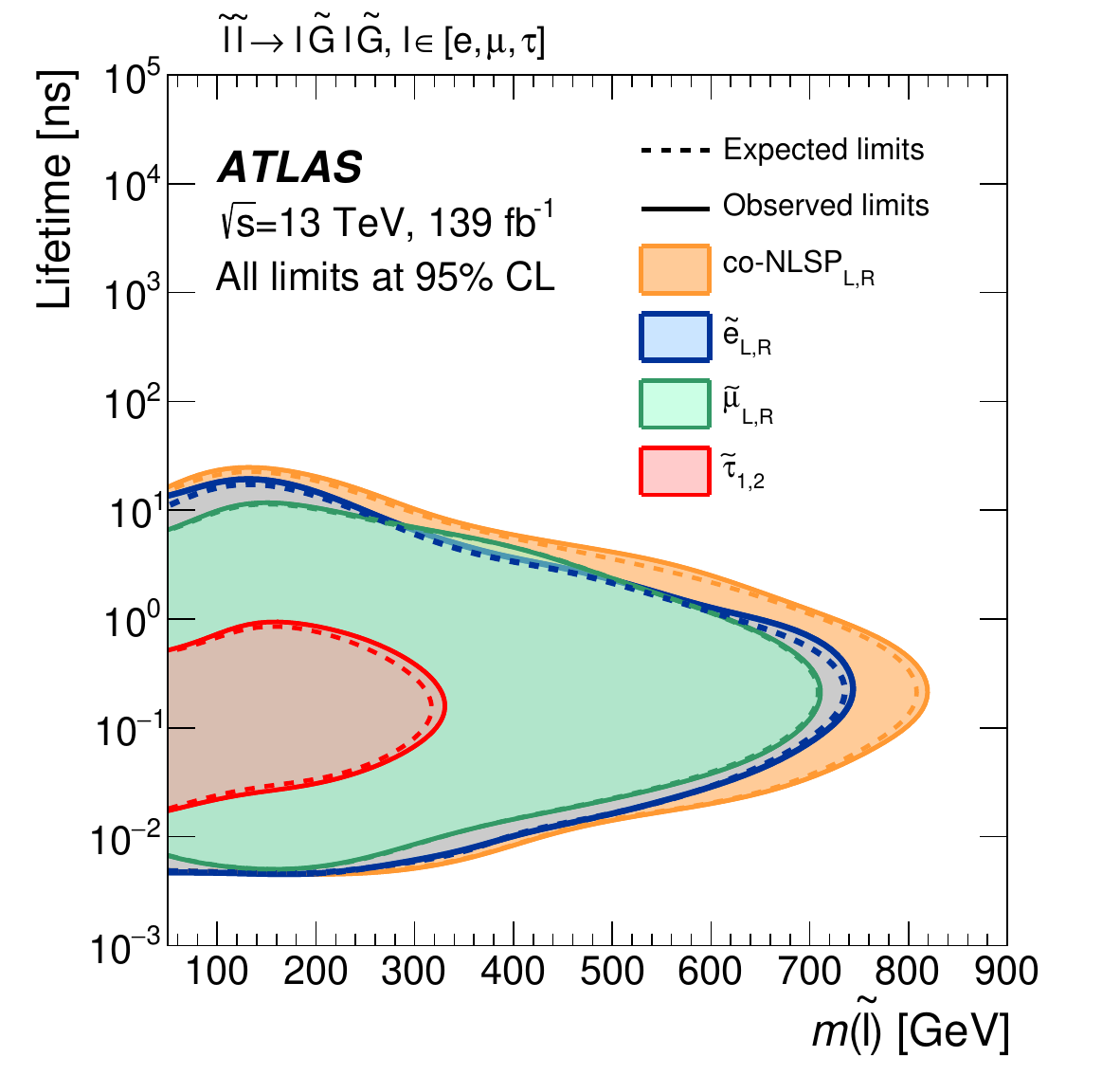}
\caption{
Expected (dashed) and observed (solid) exclusion contours for $\tilde{e}$ NLSP, $\tilde{\mu}$ NLSP, $\tilde{\tau}$ NLSP, and co-NLSP production as a function of the slepton $\tilde{\ell}$ mass at 95\% CL. Selectrons, $\tilde{e}_{L,R}$, and smuons, $\tilde{\mu}_{L,R}$, are the superpartners of the left- and right-handed electrons and muons. Staus, $\tilde{\tau}_{1,2}$ are the mixed states of the superpartners of the left- and right-handed $\tau$-leptons, with mixing angle $\sin\theta_{\tilde\tau}=0.95$. The different $\tilde{\ell}$ chiral states are assumed to be mass degenerate.
}
\label{fig:excl_limit}
\end{figure}
\FloatBarrier


We thank CERN for the very successful operation of the LHC, as well as the
support staff from our institutions without whom ATLAS could not be
operated efficiently.
 
We acknowledge the support of ANPCyT, Argentina; YerPhI, Armenia; ARC, Australia; BMWFW and FWF, Austria; ANAS, Azerbaijan; SSTC, Belarus; CNPq and FAPESP, Brazil; NSERC, NRC and CFI, Canada; CERN; ANID, Chile; CAS, MOST and NSFC, China; COLCIENCIAS, Colombia; MSMT CR, MPO CR and VSC CR, Czech Republic; DNRF and DNSRC, Denmark; IN2P3-CNRS and CEA-DRF/IRFU, France; SRNSFG, Georgia; BMBF, HGF and MPG, Germany; GSRT, Greece; RGC and Hong Kong SAR, China; ISF and Benoziyo Center, Israel; INFN, Italy; MEXT and JSPS, Japan; CNRST, Morocco; NWO, Netherlands; RCN, Norway; MNiSW and NCN, Poland; FCT, Portugal; MNE/IFA, Romania; JINR; MES of Russia and NRC KI, Russian Federation; MESTD, Serbia; MSSR, Slovakia; ARRS and MIZ\v{S}, Slovenia; DST/NRF, South Africa; MICINN, Spain; SRC and Wallenberg Foundation, Sweden; SERI, SNSF and Cantons of Bern and Geneva, Switzerland; MOST, Taiwan; TAEK, Turkey; STFC, United Kingdom; DOE and NSF, United States of America. In addition, individual groups and members have received support from BCKDF, CANARIE, Compute Canada, CRC and IVADO, Canada; Beijing Municipal Science \& Technology Commission, China; COST, ERC, ERDF, Horizon 2020 and Marie Sk{\l}odowska-Curie Actions, European Union; Investissements d'Avenir Labex, Investissements d'Avenir Idex and ANR, France; DFG and AvH Foundation, Germany; Herakleitos, Thales and Aristeia programmes co-financed by EU-ESF and the Greek NSRF, Greece; BSF-NSF and GIF, Israel; La Caixa Banking Foundation, CERCA Programme Generalitat de Catalunya and PROMETEO and GenT Programmes Generalitat Valenciana, Spain; G\"{o}ran Gustafssons Stiftelse, Sweden; The Royal Society and Leverhulme Trust, United Kingdom.
 
The crucial computing support from all WLCG partners is acknowledged gratefully, in particular from CERN, the ATLAS Tier-1 facilities at TRIUMF (Canada), NDGF (Denmark, Norway, Sweden), CC-IN2P3 (France), KIT/GridKA (Germany), INFN-CNAF (Italy), NL-T1 (Netherlands), PIC (Spain), ASGC (Taiwan), RAL (UK) and BNL (USA), the Tier-2 facilities worldwide and large non-WLCG resource providers. Major contributors of computing resources are listed in Ref.~\cite{ATL-SOFT-PUB-2020-001}.
 
 
\clearpage

\printbibliography

\clearpage 
 
\begin{flushleft}
\hypersetup{urlcolor=black}
{\Large The ATLAS Collaboration}

\bigskip

\AtlasOrcid[0000-0002-6665-4934]{G.~Aad}$^\textrm{\scriptsize 101}$,    
\AtlasOrcid[0000-0002-5888-2734]{B.~Abbott}$^\textrm{\scriptsize 128}$,    
\AtlasOrcid[0000-0002-7248-3203]{D.C.~Abbott}$^\textrm{\scriptsize 102}$,    
\AtlasOrcid[0000-0002-2788-3822]{A.~Abed~Abud}$^\textrm{\scriptsize 36}$,    
\AtlasOrcid[0000-0002-1002-1652]{K.~Abeling}$^\textrm{\scriptsize 53}$,    
\AtlasOrcid[0000-0002-2987-4006]{D.K.~Abhayasinghe}$^\textrm{\scriptsize 93}$,    
\AtlasOrcid[0000-0002-8496-9294]{S.H.~Abidi}$^\textrm{\scriptsize 29}$,    
\AtlasOrcid[0000-0002-8279-9324]{O.S.~AbouZeid}$^\textrm{\scriptsize 40}$,    
\AtlasOrcid{N.L.~Abraham}$^\textrm{\scriptsize 156}$,    
\AtlasOrcid[0000-0001-5329-6640]{H.~Abramowicz}$^\textrm{\scriptsize 161}$
% ...
\bigskip
\\

% ...
$^{29}$Physics Department, Brookhaven National Laboratory, Upton NY; United States of America.\\
% ...
$^{36}$CERN, Geneva; Switzerland.\\
% ...
$^{40}$Niels Bohr Institute, University of Copenhagen, Copenhagen; Denmark.\\
% ...
$^{53}$II. Physikalisches Institut, Georg-August-Universit\"{a}t G\"ottingen, G\"ottingen; Germany.\\
% ...
$^{93}$Department of Physics, Royal Holloway University of London, Egham; United Kingdom.\\
% ...
$^{101}$CPPM, Aix-Marseille Universit\'e, CNRS/IN2P3, Marseille; France.\\
$^{102}$Department of Physics, University of Massachusetts, Amherst MA; United States of America.\\
% ...
$^{128}$Homer L. Dodge Department of Physics and Astronomy, University of Oklahoma, Norman OK; United States of America.\\
% ...
$^{156}$Department of Physics and Astronomy, University of Sussex, Brighton; United Kingdom.\\
% ...
$^{161}$Raymond and Beverly Sackler School of Physics and Astronomy, Tel Aviv University, Tel Aviv; Israel.\\
% ...

\end{flushleft}


\end{document}